\documentclass[a4paper,11pt]{article}
\usepackage{pos}

\newcommand{\eq}[1]{\begin{equation} #1 \end{equation}}
\newcommand{\cL}{{\cal L}}
\newcommand{\cO}{{\cal O}}
\newcommand{\A}{{\cal A}}
\newcommand{\N}{{\cal N}}
\newcommand{\F}{{\cal F}}
\newcommand{\cH}{{\cal H}}
\newcommand{\Br}{{\cal B}}
\newcommand{\Fig}[1]{Figure~\ref{#1}}
\newcommand{\Tab}[1]{Table~\ref{#1}}
\newcommand{\Reff}[1]{Ref.~\cite{#1}}
\newcommand{\av}[1]{\langle #1 \rangle}

\title{Anomalies in $b\to s \ell\ell$ transitions and Global Fits}

\author{Javier Virto}

\affiliation{  
Departament de Física Quàntica i Astrofísica, Institut de Ciències del Cosmos,\\
Universitat de Barcelona, Martí Franquès 1, E08028 Barcelona, Catalunya
}

\emailAdd{jvirto@ub.edu}

\abstract{We review the status of the anomalies in $b\to s \ell\ell$ transitions, and comment on the impact of the most recent measurements in 2019 and 2020 on the global fits. We also discuss a few developments in the theory calculation of local and non-local form factors.}

\FullConference{%
  BEAUTY2020\\
  21-24 September 2020\\
  Kashiwa, Japan (online)}

\begin{document}
\maketitle

\section{Introduction}

$B$ decays mediated by the quark-level $b\to s\ell\ell$ transition are a class of semileptonic Flavor-Changing Neutral Currents (FCNCs) called \emph{rare decays} of which I will discuss two types of observables: those with $\ell=\mu$, and those measuring $\mu$-$e$ flavor non-universality. All these are very suppressed in the SM, however FCNCs are no longer ``rare'' at the LHC: only in 2016 LHCb observed more than 2000 $B \to K^*\mu\mu$ events, and they will be even more common after Phase-II LHCb upgrade~\cite{LHCbEoI} and Belle-II~\cite{Kou:2018nap}. They are still rare in the sense that they are very sensitive to NP, but they are observed in abundance, which means we have to pay a lot of attention to theory predictions and to QCD uncertainties. This is an opportunity, since this means we can use all these decays to study both New Physics (NP) and QCD simultaneously.

These decays are very suppressed in the SM because they are loop suppressed, and can have other suppressions related to CKM hierarchy and helicity. Thus the SM contributions compete potentially with tree-level BSM contributions such as those arising from massive neutral gauge bosons or leptoquarks. 

\begin{figure}
    \centering
    \includegraphics[width=15cm]{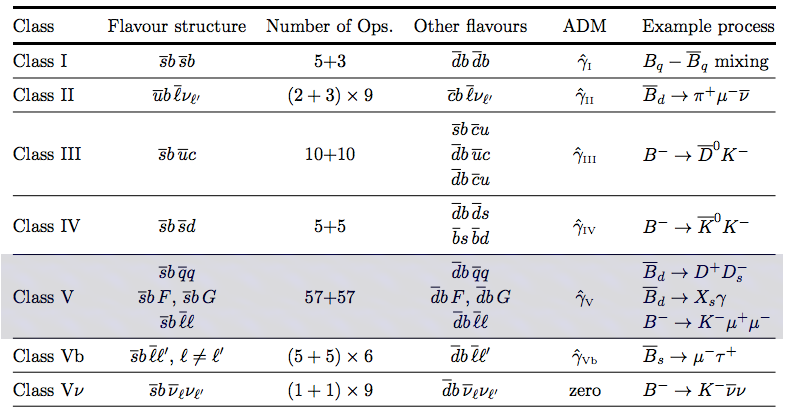}
    \caption{EFT operators relevant for $B$ physics, divided into RG-invariant classes. From~\Reff{Aebischer:2017gaw}.}
    \label{WET}
\end{figure}

$B$ mesons are stable under flavor-conserving interactions, such as QCD and QED, which means they decay (and mix) only due to weak and (possibly) BSM interactions.
In addition, $B$ mesons have a mass of about $m_B=5\,\text{GeV}\ll \Lambda_\text{EW}\sim 100\,\text{GeV}$, much smaller than the EW scale $\Lambda_\text{EW}$. If new particles are also above the EW scale, then $B$-physics as a whole must be studied in the framework of an EFT. This EFT is called the WET~\cite{Aebischer:2017gaw} (sometimes LEFT), and consists of QED and QCD complemented with a set of higher dimensional effective operators:
\eq{
\cL_\text{WET} = \cL_\text{QCD+QED} + \sum_i C_i \cO_i\ .
}
There are really many operators in this EFT. In fact there is an infinite number, but if you stop at dimension-six (as is usual and enough in $B$-physics), there are still many of them. \Fig{WET} shows all these operators classified according to their flavor quantum numbers~\cite{Aebischer:2017gaw}. Since both QCD and QED conserve flavor, operators in different classes do not mix under renormalization, and thus these classes constitute different sectors which are completely differentiated.
The relevant sector for the study of $b\to s\ell\ell$ transitions is the one called "Class V", shown in gray shading in~\Fig{WET}.
Still this class contains 114 independent effective operators, and therefore the problem of determining the coefficients of these operators from data is still a difficult one.

However, most of these operators are four-quark operators, and in the NP analysis of $b\to s\ell\ell$ transitions we are concentrating on a much smaller set: the semileptonic vector and axial-vector operators $\cO^{(\prime)}_{9\ell},\cO^{(\prime)}_{10\ell}$, the dipole operators $\cO^{(\prime)}_{7}$ and the four-quark ``current-current'' operators with charm $\cO_{1,2}$, which are important for QCD corrections.  It is rather clear by now that this set is the right one to focus on.
The idea now is to use all the available data to constrain potential NP contributions to the semileptonic and dipole operators.
{\bf Currently, global determinations of $C_{9\mu}$ (and maybe $C_{10\mu}$) seem discrepant with SM predictions, with an important statistical significance.}
The purpose of this talk is to give a summary and update of the current situation with regard to these global fits.

\section{How did we get here? A historical digression}

Fits to $b\to s$ transitions are nothing new, and have been done for more than 26 years~\cite{Ali:1994bf}. However some key measurements such as $B_s\to \mu^+\mu^-$ or $B\to K^{(*)}\mu\mu$ awaited LHCb, and flavor physicists prepared for that.
In 2001 there was the paper by Beneke, Feldmann and Seidel~\cite{Beneke:2001at}, where they applied the (recently developed)  QCD-facorization machinery to $B\to K^*\mu^+\mu^-$.
They calculated theory predictions at the kinematic region of low $q^2$ (invariant dilepton mass). One thing they found is that the zero crossing of the Forward-Backward asymmetry (denoted by $q_0$) at leading order in $\alpha_s$ satisfies the following equation:
\eq{
C_9 + \text{Re}(Y(q_0^2)) = -\frac{2M_B m_b}{q_0^2} C_7^\text{eff}
}
which is free from hadronic uncertainties at this order in the QCD-factorization expansion. They also calculated $\alpha_s$ corrections, and showed that indeed hadronic uncertainties are significantly reduced for this observable. This was the first ``clean'' observable in $B\to K^*\mu^+\mu^-$.
The reason for this cancellation of hadronic uncertainties is that out of the total of seven ``local'' form factors that enter the amplitudes, in this approximation ($E_{K^*}\sim \Lambda_\text{QCD}\ll m_b$) there are only two independent combinations~\cite{Charles:1998dr,Beneke:2000wa,Bauer:2000yr}, and so it is possible to build observables (such as $q_0$) where form factors cancel out but still a good sensitivity to the Wilson coefficients is maintained.

The question now was, can one build observables where this happens, not only at a single value of $q^2$, but as \emph{functions of $q^2$}? This issue was studied by Kruger and Matias~\cite{Kruger:2005ep}, which came up with the observable $A_T^{(2)}$, which is very clean in the SM but also very sensitive to right-handed currents.
A few people worked along these lines thereafter, and it was about a decade later that a complete \emph{basis} for \emph{all} observables of this type was found~\cite{Matias:2012xw,DescotesGenon:2012zf}, among which $P_5'$ is. These observables were called ``optimized" or ``form-factor independent". A similar game can be played in the region of large $q^2$~\cite{Bobeth:2012vn,Descotes-Genon:2013vna}.

Shortly after (April and August 2013), LHCb published two papers~\cite{Aaij:2013iag,Aaij:2013qta}, with a measurement of the full basis of observables. It was the second paper that contained the measurement of $P_5'$, and uncovered the ``$P_5'$ anomaly"~\cite{nicoserraEPS}. We immediately run a global fit including all this data and found a good fit for $C_{9\mu}^\text{NP}\simeq -1$ ($-25\%$ NP contribution compared to the SM), and the SM outside the $\sim 4\,\sigma$ confidence-level region\cite{Descotes-Genon:2013wba}.
Most importantly, {\bf this anomaly was not only  driven by $P_5'$, and was of a more global nature}, motivating the broader name ``$B\to K^*\mu\mu$ Anomaly''.
Soon after a few other papers came out confirming these findings~\cite{Altmannshofer:2013foa,Beaujean:2013soa,Horgan:2013pva}. The most striking one was the one from the lattice group, where they used newly computed form factors from Lattice QCD on the large-$q^2$ region and whose fit included only observables at low recoil, hence a rather independent confirmation.

\begin{table}
\centering
\setlength{\tabcolsep}{14pt}
\renewcommand{\arraystretch}{1.3}
\begin{tabular}{@{}lccc@{}}
\hline
Observable & Experiment & SM prediction  & pull \\
\hline
$R_K^{[1.1,6]}$ & $0.85\pm 0.06$ & $1.00\pm 0.01$ & $+2.5\sigma$ \\ 
$R_{K^*}^{[0.045,1.1]}$ & $0.66^{+0.11}_{-0.07}$ & $0.92\pm0.02$ & $+2.3\sigma$ \\
$R_{K^*}^{[1.1,6]}$ & $0.69^{+0.12}_{-0.08}$ & $1.00\pm 0.01$ &  $+2.6\sigma$ \\
\hline
$\av{P_5'}_{[4,6]}$ & $-0.44\pm0.12$ & $-0.82\pm 0.08$ &  $-2.7\sigma$ \\
$\av{P_5'}_{[6,8]}$  &  $-0.58\pm0.09$ & $-0.94\pm 0.08$ & $-2.9\sigma$ \\
$\Br_{\phi\mu\mu}^{[2,5]}$ & $0.77\pm0.14$ &   $1.55\pm0.33$ & $+2.2\sigma$ \\
$\Br_{\phi\mu\mu}^{[5.8]}$ & $0.96\pm0.15$  & $1.88\pm 0.89$ & $+2.2\sigma$ \\
\hline
\end{tabular}
\caption{List of main ``anomalous'' measurements as of 2020.}
\label{TableAnom}
\end{table}

The discovery of the $B\to K^*\mu\mu$ anomaly was followed by a year characterized by a mixture of excitement and ``warm'' discussions on hadronic uncertainties. Then, in the summer of 2014 there was a new \emph{twist}: the measurement of $R_K$ by LHCb~\cite{Aaij:2014ora}:
\eq{
R_K\equiv \frac{{\cal B}(B^+\to K^+\mu\mu)_{[1,6]\text{GeV}^2}}{{\cal B}(B^+\to K^+ee)_{[1,6]\text{GeV}^2}};
\quad
R_K^\text{SM} = 1;
\quad
R_K^\text{LHCb\,2014} \simeq 0.75\pm 0.1\ ,
}
off by short of $3\,\sigma$ from the very precise SM prediction~\cite{Hiller:2003js,Bobeth:2007dw,Bordone:2016gaq}.
This was quite unexpected at the time, as it required more ``agressive'' New Physics, violating Lepton Flavor Universality. The crucial observation was the one put forward originally in~\cite{Alonso:2014csa} (see also~\cite{Hiller:2014yaa,Ghosh:2014awa}), which was that if you assume that NP contributes only to the muonic operators, then the same New Physics explains simultaneously both the $B\to K^*\mu\mu$ and $R_K$ anomalies.
This was very important because it made the anomaly coherent in several different fronts.
This feature, that the $b\to s\mu\mu$ anomalies explain the LFNU anomalies, survived many years. There were updates on the $B\to K^*\mu\mu$ angular distribution, new measurements of $R_{K^*}$, measurements of $B\to K^*\ell\ell$ by Belle distinguishing muon and electron observables, new measurements of $B\to K^*\mu\mu$ by ATLAS and CMS, measurements of angular observables in $B_s\to \phi\mu\mu$, etc., but updates of the global fits kept confirming this ``economical'' consistency between $b\to s\mu\mu$ and LFNU~\cite{Descotes-Genon:2015uva,Celis:2017doq,Capdevila:2017bsm}.
This seems no longer the case after the new 2019-2020 measurements of $P_5'$ and $R_K$, but it makes a lot of sense when the coupling to the third family of leptons is large~\cite{Crivellin:2018yvo,Alguero:2019ptt}.
The year 2014 was also when the link to the anomalies in semileptonic $b\to c \ell\nu$ was put forward~\cite{Bhattacharya:2014wla}. I will not discuss this here. The implications of this link between charge and neutral LFUV currents has been discussed 
by Gino Isidori in this conference.

\section{Current status of available measurements as of 2020}

There are currently a total of 180 observables going into the global fit: many different modes ($B_s\to \mu\mu$, $B\to X_s\mu\mu$, $B\to K^*\gamma$, $B\to X_s\gamma$, $B\to K\mu\mu$, $B\to K^*\mu\mu$, $B_s\to \phi\mu\mu$), including inclusive and exclusive, different types of observables (branching ratios and angular observables), at different regions of $q^2$ (low and high, with different theory approaches), observables probing lepton universality ($R_K$, $R_{K^*}, Q_{4,5}$~\cite{Capdevila:2016ivx}), and from five different experimental collaborations: LHCb, Belle, Babar, ATLAS and CMS.
This is a very wide range of different inputs to the fit.
The latest updates~\cite{Alguero:2019ptt}, which I will be summarizing here, include the 2019 measurements of $R_K$ by LHCb~\cite{Aaij:2019wad} and Belle~\cite{Choudhury:2019imr}, and the new 2020 $B\to K^*\mu\mu$ analysis by LHCb with all Run 1 + 2016 data~\cite{Aaij:2020nrf}.
There is also an LHCb analysis of $B^+\to K^{*+}\mu\mu$~\cite{Aaij:2020ruw} that I will not consider, but which has been included in the fits in~\cite{Ciuchini:2020gvn,Hurth:2020ehu}.

The list of anomalies (\Tab{TableAnom}) containing the measurements deviating more than $2\,\sigma$ has been essentially unchanged for some years now. Recent updates have moved the various pulls slightly, but qualitatively the picture is the same. The important observation here is that is not only the anomalies one must worry about; when doing a global fit is also important not to deviate from the measurements that agree with the SM. The global fit is in this sense the right thing to do, to find the consistency between all measurements, the ones that deviate and the ones that don't.

\begin{figure}
    \centering
    \includegraphics[width=15cm]{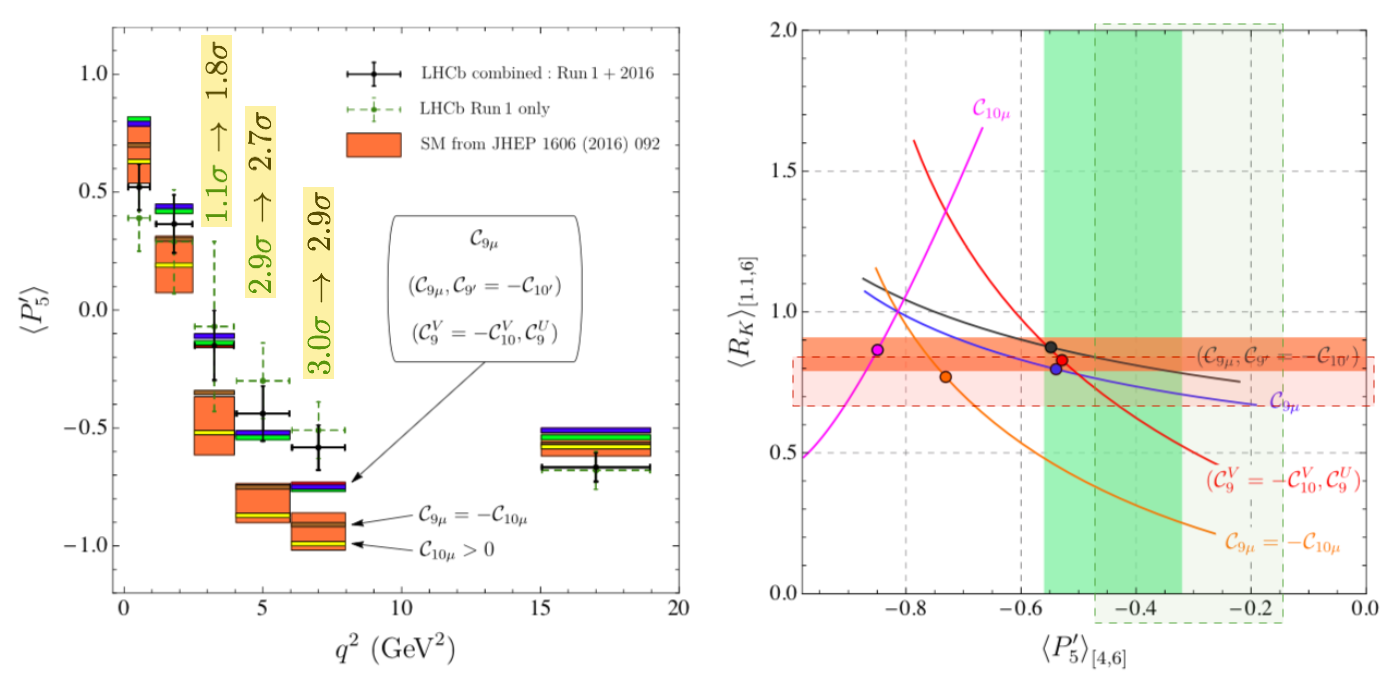}
    \caption{Some details on the new measurements of $P'_5$ and $R_K$ (mostly from~\Reff{Alguero:2019ptt}).}
    \label{newP5pRK}
\end{figure}

Concerning the new measurements of $R_K$ and $P_5'$. \Fig{newP5pRK} (left) shows $P'_5$, comparing previous and new measurements. The central values in the three relevant bins have moved towards the SM but the errors have diminished, and so the significances are mostly unchanged. That the measurements approach the SM is not necessarily bad for NP. One can see that the predictions from the most relevant NP benchmarks favored in the global fit are and have been below the data. Thus, now that the data has moved downwards, the measurements of $P_5'$ are more consistent with the NP scenarios that fit well the rest of the data.
The same is true for the updated measurement of $R_K$, as can be seen in the right panel of~\Fig{newP5pRK} . {\bf Both $R_K$ and $P_5'$ have moved towards the SM, but this actually improves the NP fits.} This is something to keep in mind when judging future measurements.

\section{Some details on theory predictions}

The main inputs to the fits are the $B\to M\ell\ell$ observables. The corresponding amplitudes (to all orders in QCD) are given by the formula~\cite{Bobeth:2017vxj}
\eq{
\A_\lambda^{L,R} = \N_\lambda\ \bigg\{
(C_9 \mp C_{10}) {\color{blue} \F_\lambda(q^2)}
+\frac{2m_b M_B}{q^2} \bigg[ C_7 {\color{blue} \F_\lambda^{T}(q^2)}
- 16\pi^2 \frac{M_B}{m_b} {\color{red} \cH_\lambda(q^2)} \bigg]
\bigg\} + \cO(\alpha_\text{em}^2) \ .
}
There are two things one has to worry about from the theory point of view: (1) the local form factors (in blue) defined as matrix elements of local currents, $\F_\lambda^{(T)}(q^2) \sim \langle\bar M_\lambda(k)| \,\bar s\, \Gamma^{(T)}_\lambda\, b\, |\bar B(k+q)\rangle$, and (2) the non-local form factors (in red), which are more complicated matrix elements of a non-local operator:
$\cH_\lambda(q^2) \sim i \,{\cal P}_\mu^\lambda \int d^4 x\ e^{i q\cdot x}\,
\langle \bar M_\lambda(k)|
T\big\{ {\cal J}_{\rm em}^\mu(x), C_i \,\cO_i(0) \big\} | \bar B(q+k)\rangle$.

The local form factors can be calculated within lattice QCD at high $q^2$ (e.g.~\cite{Horgan:2013hoa}) and within two independent types of light-cone sum rules at low $q^2$~\cite{Straub:2015ica,Descotes-Genon:2019bud}.
One can fit simultaneously the large- and low-$q^2$ determinations, because we know the $q^2$ dependence, since this is given by the analyticity of the matrix elements (so this is theoretically solid).
Everything seems to be consistent~\cite{Straub:2015ica,Gubernari:2018wyi}, and the error bars are at the level of $10-15\%$. We are of course eager to reduce these uncertainties but this is not the most pressing issue in the theory predictions at this time.

Concerning the non-local form factors, there is mostly a consensus about the strategy to follow~\cite{Khodjamirian:2012rm,Asatrian:2019kbk,Gubernari:2020eft,Beylich:2011aq}. One can use different types of OPEs to calculate the non-local form factors at negative $q^2$ (outside the physical region) or at large positive $q^2$. For the region of low-$q^2$, since data lives at positive $q^2$, one needs to analytically continue the calculation at negative $q^2$ to the physical region.
One of the main references is Ref.\cite{Khodjamirian:2010vf}, which uses a dispersion relation and data to perform this extrapolation. This is what we are currently using in the global fit, but there are plans to improve on this (see below).

The calculation of non-local contributions can also be checked {\it a posteriori} from the $B\to M\ell\ell$ data. For instance one can perform fits to NP Wilson coefficients grouping the data in classes within specific $q^2$ bins. If the calculated non-local contributions that are an input to the fit are correct, all these fits should give compatible results~\cite{Descotes-Genon:2015uva}. So far, this is the case~\cite{Alguero:2019ptt}.

\section{Updated fits}

The global fits as of 2019 are discussed in~\cite{Alguero:2019ptt} (see also~\cite{Aebischer:2019mlg,Alok:2019ufo}). The results include fits to one, two and six independent combinations of (NP contributions to) Wilson coefficients. Fits including different sets of observables are considered, most importantly the fit to the complete set of 180 $b\to s \ell\ell$ observables (the ``all'' fit) as well as the fit to (only) all observables that measure $e-\mu$ non-universality (the ``LFNU'' fit). But also separate fits to LHCb/Belle/ATLAS/CMS data are considered in order to check the consistency of the data provided by the different experimental collaborations. 

The new measurements in 2019 and 2020 do not change the picture but provide refinements. Significant one- and two-dimensional hypotheses all involve NP in $C_{9\mu}$, and feature a $p$-value of around $30-50\%$ ($60-90\%$), and a SM pull of around $6\,\sigma$ ($3.5\,\sigma$) for the all (LFNU) fits. In the case of the six-dimensional (all) fit, the $p$-value is around $50\%$ and the SM pull is $5.8\,\sigma$, and $C_{9\mu}$ is different from its SM value at more than $3\,\sigma$. In general, the new fits show smaller $p$-values, but this is due to the smaller experimental uncertainties in the new $B\to K^*\mu\mu$ data.

\section{A few improvements for the future}

A few developments related to theory predictions for local and non-local form factors have been presented in~\cite{Descotes-Genon:2019bud,Bobeth:2017vxj,Gubernari:2020eft}, and have not been yet implemented in the global fits. The first one has to do with the finite-width effects from the fact that the $K^*$ is a strong resonance that decays into a $K\pi$ pair~\cite{Descotes-Genon:2019bud}. This issue affects the local and non-local form factors in the amplitude, and can be addressed by using a generalization of the method of $B$-meson Light-Cone Sum Rules~\cite{Cheng:2017smj}. The local  $B\to K\pi$ form factors and their impact on $B\to K^*\ell\ell$ are discussed in great generality in~\Reff{Descotes-Genon:2019bud}. The width-to-mass ratio of the $K^*(892)$ is about $5\%$, however it is found that the finite-width correction in the local form factors is twice that number. In addition, this is approximately a universal effect in all form factors, which means that (1) the theory prediction for the branching fraction is increased by about $20\%$, and (2) observables that are defined as ratios (such as $P'_5$) are mostly unaffected. The method also allows to gauge the interference in the predictions from higher resonances such as $K^*(1410)$ or $K^*_0(1430)$, and to confront them with measurements in this region~\cite{Aaij:2016kqt}.

The second improvement has to do with the non-local form factors. A recalculation of the subleading effects discussed in~\cite{Khodjamirian:2010vf} has been performed in~\Reff{Gubernari:2020eft}, including new ingredients and updates. The result is surprising: two orders of magnitude smaller at negative $q^2$. This renders the subleading corrections negligible for current purposes, meaning that using the OPE calculation at leading order is sufficient. In addition to the theory calculation at negative $q^2$, the continuation to the physical region has been also revisited in~\cite{Bobeth:2017vxj,Gubernari:2020eft}. The result is a parametrization for the $q^2$ dependence that can be fixed by theory data at negative $q^2$ combined with measurements of $B\to \psi_n K^*$ at the $J/\psi$ and $\psi(2s)$ poles~\cite{Bobeth:2017vxj}, and constrained by a unitarity bound~\cite{Gubernari:2020eft}. Details can be found in these references, where also the extension to $B_s\to \phi\ell\ell$ is given. Global fits including and studying all these improvements are in progress.

\section{Summary}

The set of $b\to s\ell\ell$ anomalies are alive and a global coherence remains after the measurements in 2019 and 2020. While some key measurements have central values that have moved towards the SM predictions, the overall consistency of the data and the global tension with respect to the SM has not diminished (in fact this may even enhance the significance for NP). The NP fit is a good fit (with $p$-values in the ballpark of $50\%$), and the SM pull is high (around $6\,\sigma$ and $3.5\,\sigma$ for the ``all'' and ``LFNU'' fits respectively). In addition, data sets from all different experiments are compatible (although still dominated decisively by LHCb).

On the QCD side (less popular but terribly important), several developments have been made concerning local and non-local form factors, which must now be implemented in global fits.
An important point is that these developments drift towards a more ``data-driven'' philosophy, which will allow to use the experimental capabilities of the LHCb Phase-II~\cite{LHCbEoI} upgrade and Belle-II~\cite{Kou:2018nap} to improve the precision and reliability of SM predictions greatly.

\section*{Acknowledgements}

\noindent I thank the organizers and conveners of the BEAUTY'2020 conference for inviting me to give this plenary talk and for keeping the conference alive during these difficult times of sanitary confinement. There is no surrender to Beauty.
My research is funded (not too generously, but still appreciated) by the Spanish MINECO through the “Ram\'on y Cajal” program RYC-2017-21870.


\end{document}